\documentclass[fleqn,12pt,twoside]{article} 
\usepackage{espcrc1}

\usepackage{graphicx}
\usepackage{floatflt}

\newcommand{\AmS}{{\protect\the\textfont2
  A\kern-.1667em\lower.5ex\hbox{M}\kern-.125emS}}

\title{High Transverse Momentum Results from the STAR Collaboration}

        \author{G. J. Kunde
                \address{Physics Department,Yale University,\\
                  P.O. Box 208124, New Haven CT, 06520, USA}
      for the STAR Collaboration
      and the STAR-RICH Collaboration \footnote{For the full author list
      and acknowledgements see Appendix "Collaborations" in this volume.}
               }               
\begin{document}

\maketitle

\begin{abstract}
The STAR collaboration presents new measurements of high $p_T$ hadron production in Au+Au and p+p collisions at RHIC. We extend the previously
reported suppression of inclusive hadron\-s and large azimuthal anisotropies
to much higher transverse momentum, decisively establishing the existence of
strong medium effects on hadron production well into the perturbative regime.
Near-angle two-particle correlations show directly that hadron\-s at $p_T>4$
GeV/c result from the fragmentation of jets. Additional evidence for the
onset of perturbative QCD in this region comes from the flavor dependence
of the inclusive yields and elliptic flow. Finally, comparison of back-to-back
hadron pairs at high $p_T$ in p+p and in Au+Au collisions at
various centralities
reveals a striking suppression of high $p_T$ back-to-back pairs in the most
central Au+Au collisions. All of these phenomena suggest a picture in which
partons or their hadronic fragments are strongly absorbed in the
bulk matter, with the observed hadrons resulting from jets produced on the
periphery of the collision zone and directed outwards.
\end{abstract}

\section{Introduction}
At energy densities larger than $~1~$GeV/fm$^3$, matter is expected to undergo
a phase transition from a hadronic phase to a Quark Gluon Plasma \cite{qgp}.
Collisions of heavy nuclei at RHIC generate an extended volume of 
high energy density. If this density exceeds that required for a
phase transition, the evolution of the collision cannot be described
purely in terms of a hadron gas and
the underlying quark and gluon degrees of freedom must be invoked \cite{satz}.
Hard scattering of incoming partons generates high $E_T$ partons in the final
state that
propagate through the medium, eventually fragmenting into a correlated spray
of hadrons called a jet. The high $E_T$ partons interact with the
surrounding medium radiating soft gluons at a rate proportional to the
energy density of the medium. Measurement of jet energy loss may, therefore,
provide a direct probe of the energy density \cite{baier}. This talk
discusses several observables related to energy loss.

A jet may fragment in such a way that one hadron carries a large fraction of
its momentum (leading hadron). The suppression of the yield of leading
hadrons may signal jet energy loss. Strong suppression of high $p_T$
hadrons has been observed in central Au+Au collisions
\cite{phenixhighpt,starhighpt} suggesting the existence of strong in-medium
interactions, though the
contribution of partonic energy loss is masked by other effects, such as
shadowing and initial state multiple scattering (Cronin effect)
\cite{vitevspslhc}. Additional evidence for strong medium effects comes from
the observation of large elliptic flow in non-central collisions in which
the high $p_T$ hadron is strongly correlated with the orientation of the
reaction plane of the event \cite{v2charged,v2identified,v2highpt,kirill,phenixv2}.

\begin{floatingfigure}{90mm}
\includegraphics*[width=90mm]{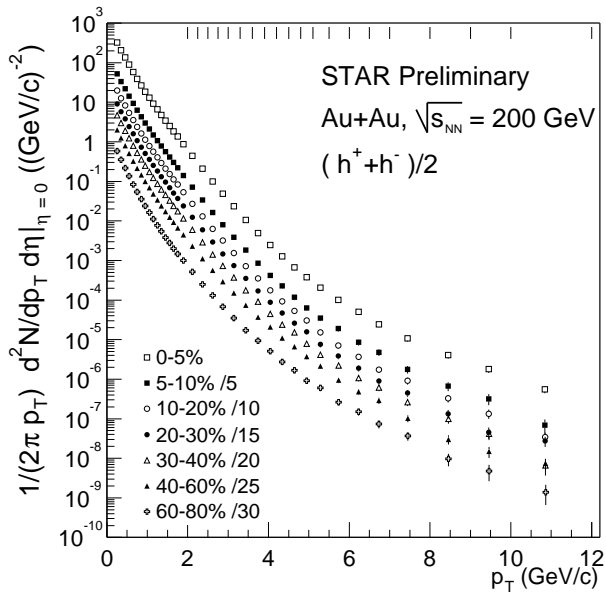}
\vspace{-2mm}
\caption{Invariant cross sections for charged hadrons
from Au+Au  collisions at $\sqrt{s_{NN}}$ = 200 GeV.
The hash marks indicate the bin limits.}
\vspace{2mm}
\label{fig:Spectra}
\end{floatingfigure}

\noindent
Neither of the observables listed above proves experimentally that the observed
hadron\-s are indeed the fragments of jets; for this we utilize
two-particle correlations. Near-angle azi\-mu\-thal
two-particle  correlations for Au+Au and p+p collisions
are in quantitative a\-gree\-ment \cite{dave}, indicating that hadron\-s
with $p_T>4$ GeV/c stem from the fragmentation of jets. This
study also shows that the fragmentation of the
{\it observed} jets is not substantially modified
in Au+Au collisions. Further evidence for the onset of pQCD is
the breakdown of the hydrodynamic description of the azimuthal anisotropy
$v_2$ above $p_T>2-3$ GeV/c, for both baryons and meson\-s. Also within
the same momentum range, the measured ratio of anti-pro\-tons to protons
begins to decrease with increasing momentum. All of these signatures
are expected when the  hadron\-ic cross sections at high transverse momentum
are dominated by pQCD processes.

The new observable, which directly probes the opacity of the medium, is the
comparison of back-to-back azimuthal correlations for Au+Au and p+p.
While a small suppression is observed in peripheral Au+Au collisions, central
collisions are nearly devoid of back-to-back correlations. The partons or
the hadrons from parton fragmentation that traverse the
center of the reaction zone are either suppressed in their yield or lose the
memory of their production (i.e. multiple scatter).


A large number of qualitatively new observations are presented for
the first time at this conference:
single inclusive spectra and $v_2$
to  $p_T<12$ GeV/c ;
the two-particle azimuthal correlations; the precision study of non-flow contributions to high $p_t$;
the meson/baryon $v_2$;
and identified particle ratios up to 4.5 GeV/c.
A full description of the physics of high $p_T$ hadron production
must accommodate all of these phenomena.


\section{Data Collection and Analysis}

This paper reports the results from data taken
with the STAR detector at $\sqrt{s_{NN}}$ = 200 GeV
during the 2001 RHIC run.  The Au+Au analysis
uses 2.7M minimum bias and 2.1M central
triggered events (10~\% of the total inelastic Au+Au cross section).
Event centrality classes and numbers of
participants were determined using the primary charged particle
multiplicity within $|\eta| <$ 0.5 \cite{gene}. The p+p data sample used
in the two-particle azimuthal correlation analysis consists of 10M minimum
bias events triggered by scintillator counters which cover $3.5 < |\eta| <
5.0$.
Comparison is also made to
the Au+Au data at $\sqrt{s_{NN}}$ = 130 GeV recorded in the first year of
physics running.

Charged particle tracks are measured in the TPC,  which was operated
in a 0.5 Tesla magnetic field.  
The momentum resolution is $\approx$ 3 \% at 5 GeV/c,
a factor of 3 better than in year 2000, in which the field was
half the strength. Particle identification
is carried out at high transverse momentum
via secondary weak decays  in the TPC \cite{gene} and the
STAR-RICH \cite{brian}
detector for primary particles at $p_T >$  1.5 GeV.

\vspace{5mm}
\section{Results}

\begin{floatingfigure}{90mm}
\includegraphics*[width=90mm]{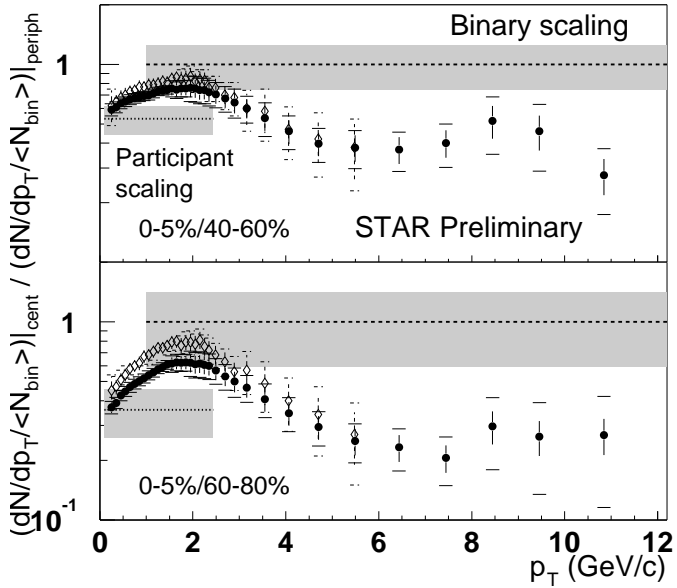}
\vspace{-2mm}
\caption{Ratio $R_{CP}$ 
as a function of transverse
momentum for central and peripheral events. The upper panel shows the ratio
of the centrality  classes  0-5\%/40-60\%, the lower panel shows
0-5\%/60-80\%. Error bars and bands are described in the text.}

\vspace{2mm}
\label{fig:rcp}
\end{floatingfigure}

The invariant yields for charged ha\-drons $(h^-+h^+)/2$ at
mid-rapidity are shown in  Fig.~\ref{fig:Spectra}. The top curve
represents the most central event selection, while all other centralities
are scaled for clarity.
The spectra are corrected for efficiency, background
and momentum resolution \cite{jenn}.
The errors indicated include both  the statistical and systematic
uncertainties, and  extend beyond the symbol size only at high momentum.

The onset of suppression of high transverse momentum hadrons in
central collisions Au+Au was recently established in the
literature \cite{phenixhighpt,starhighpt}, however 
measured in a momentum range ($p_T < 6$ GeV/c)
where soft processes still contribute.
It is essential that these measurements be extended to the highest
possible momenta, in order  to maximize the hard scattering contribution.
Two related measurements have been discussed in the past: $R_{AA}$,
which is the ratio of AA yields to the binary collision
scaled nucleon-nucleon (NN) reference, and $R_{CP}$, the
ratio of central and peripheral yields normalized by
the number of binary collisions ($\langle N_{bin} \rangle$).

At present, the NN reference is limited
to $p_T < 6$ GeV/c and comes from the UA1 experiment \cite{UA1}.
$R_{AA}$ at 130 GeV is reported in \cite{starhighpt}, while $R_{AA}$ at
200 GeV gives similar results. Here we
focus on the measurement which reaches farthest into the pQCD
domain, $R_{CP}$ for $p_T <$ 12 GeV/c. Calculations of
the number of binary collisions 
$\langle N_{bin} \rangle $ and  number of participants
$\langle N_{part} \rangle $
are discussed in \cite{starhighpt}.


Figure \ref{fig:rcp} shows the ratio of charged hadron
yields in central normalized to peripheral  collisions,
scaled by $\langle N_{bin}
\rangle $. The 200 GeV data are represented by solid circles and the
previously published 130 GeV data \cite{starhighpt} are shown as open
symbols with dashed error bars. The horizontal dashed lines and shaded areas indicate
the expectations for scaling with $\langle N_{bin}\rangle $ and
$\langle N_{part} \rangle $, together with
their respective uncertainties.

The ratio scales approximately with $\langle N_{part}
\rangle $ at low transverse momentum, and increases towards binary collision
scaling at $p_T \approx $2 GeV/c.  Beyond this, the ratio
decreases to  0.23\%+-0.06\%  at 6.4 GeV/c for the more
peripheral ratio.
The qualitatively new feature of these data is the saturation
of the suppression 
out to a $p_T$ of 12 GeV/c. The hadron production at high transverse momentum
does not follow a binary scaling but is roughly consistent with the participant
scaling (\cite{roland}, this proceedings).
This will be discussed in more detail at the end of the paper.


\begin{floatingfigure}{90mm}
\includegraphics*[width=87mm]{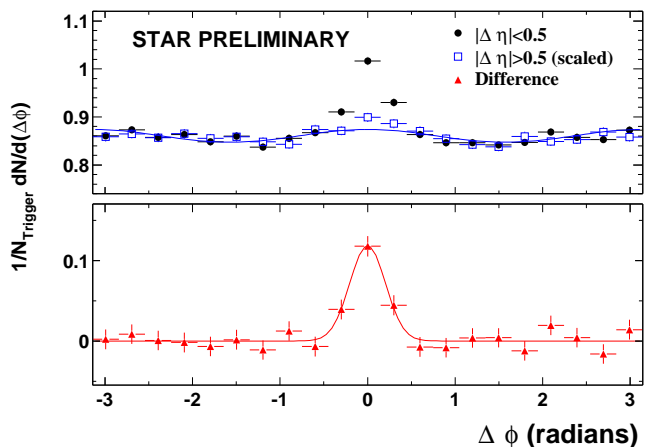}
\vspace{-6mm}
\caption{Azimuthal two-particle correlations for central Au+Au collisions
  with a trigger and associate particle for two intervals
  in relative pseudo-rapidity. The bottom figure shows the
  difference between the
  two measurements, suggestive of hard scattering. For details see
  \cite{dave}.}
\vspace{3mm}
\label{fig:azimuthal}
\end{floatingfigure}

\noindent
In order to address directly the origin of the high $p_T$ hadrons, we
search for jet-like correlations using two-par\-ticle azimuthal
correlations, see also \cite{v2highpt}. From nucleon-nucleon collisions,
it is known that the leading particles in a  jet cone have a width in
pseudo-rapidity of $\Delta \eta \approx$ 0.5. In the
Au+Au case, in addition to the jet contribution one expects azimuthal
correlations due to elliptic flow, which 
are not localized in pseudo-rapidity. To extract the jet-specific
correlations, we construct two azimuthal correlation functions
of high transverse momentum particles
for the relative pseu\-do-rapidity intervals
of $\Delta \eta<$0.5 and $\Delta \eta>0.5$.

Figure \ref{fig:azimuthal} shows the two-particle correlation function
for a trigger particle in the range  4 $< p_T <$ 6 GeV/c and
all associated particles in that event from 2 GeV/c
$< p_T^{associated} < p_T^{trigger}$.
The solid and open symbols represent
the small and large pseudo-rapidity differences, respectively.
The lower panel shows the difference between the two
measurements which should not contain any flow effects but only
near-side hard scattering and resonance contributions.
(Note that due to the subtraction all back-to-back contributions
cancel). The peak at small angular difference is pronounced and
is suggestive of jets.
Additional studies  lead to the conclusion that the cross section
for charged hadron production
above 4 GeV/c is largely  due 
to hard scattering processes.
For a more detailed discussion see \cite{v2highpt,dave}.

In a pQCD model with partonic energy loss, a strong coupling of momentum and
coordinate space is predicted due to high gluon densities. This effect
may be observable in non-central collisions. Hard scattered
partons undergo energy loss, whose magnitude depends on the path length of the
traversed medium: initial spatial anisotropies are therefore translated
into momentum anisotropies.
At lower $p_T$, elliptic flow \cite{v2charged,v2identified}
dominates the measurements.
The asymmetry is quantified using the observable $v_2$,
the second Fourier component of the
azimuthal distribution of high $p_T$ charged tracks with respect to the
reaction plane. The reaction plane in turn
is determined by low transverse momentum particles with $p_T<2$ GeV/c.
The detailed presentation can be found in \cite{kirill} and references
therein. In this paper only two highlights will be
discussed, the $v_2$ at high transverse momentum  $p_T < 12$ GeV/c
and the particle-identified $v_2$.

\begin{floatingfigure}{90mm}
\includegraphics*[width=85mm]{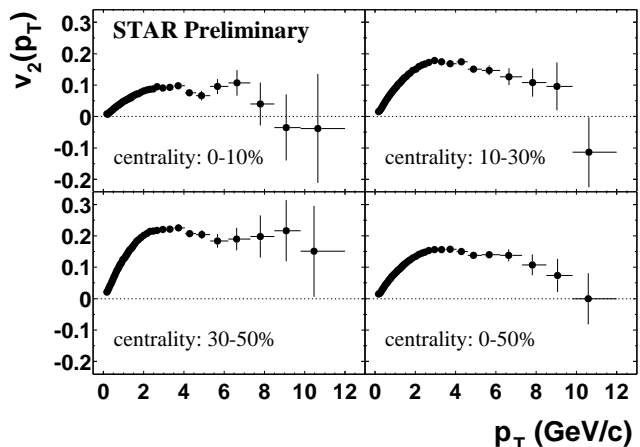}
\vspace{-5mm}
\caption{Azimuthal anisotropy $v_2$ for charged
hadrons as a function of transverse momentum up to 12 GeV/c
for 3 different centrality classes and for minimum-bias events.
Only statistical errors are shown.}
\vspace{0mm}
\label{fig:12gev}
\end{floatingfigure}

Figure \ref{fig:12gev} shows 
$v_2$ for charged ha\-dron\-s as a function of the transverse momentum
up to 12 GeV/c for  most central events (0-10\% ), followed by two centrality
classes (10-30\% and 30-50\%) and minimum-bias events.
The common feature of all three
centrality classes is the rising $v_2$,
which saturates around 2-4 GeV/c. For non-central collisions, the observed
an\-iso\-tropy is finite out to the highest measured transverse momentum,
well into the pQCD regime. The persistence of finite anisotropy
at these momenta makes the interpretation in terms of partonic or
leading hadron energy loss in an asymmetric geometry rather compelling.

The validation of hydrodynamic calculations in predicting the
elliptic flow in the soft regime lies in the particle identified
measurement. At high transverse momentum, recent theoretical papers
predict a different evolution of the anisotropy for mesons and baryons
as a function of their transverse momentum \cite{glvp,flowko}.
Junction transport and
hydrodynamics may have a large baryon cross section at intermediate
$p_T$, while pions may be dominated by pQCD production mechanisms starting
at a lower $p_T$.

\begin{figure}
\includegraphics*[width=80mm]{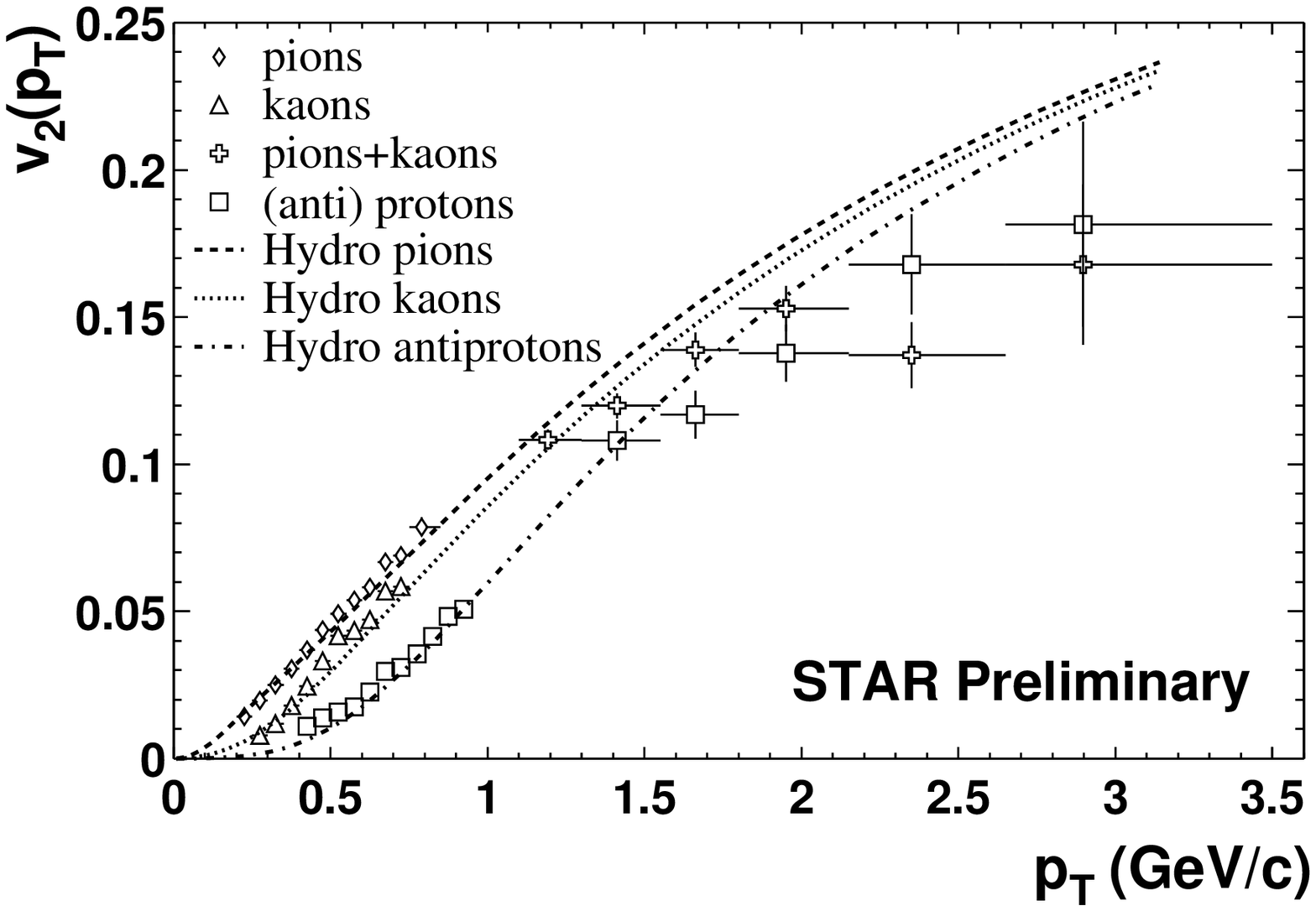}
\includegraphics*[width=80mm]{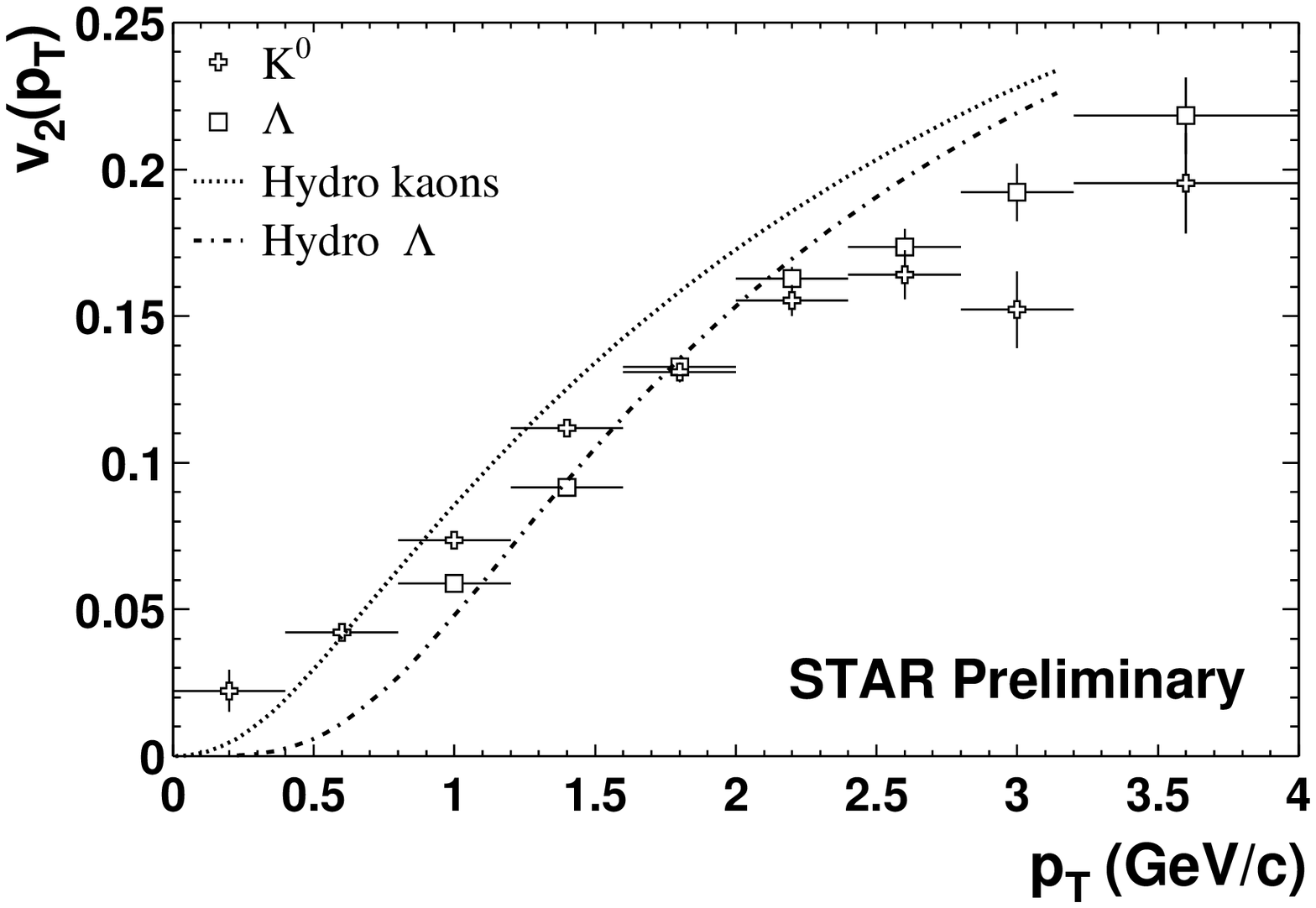}
\vspace{-10mm}
\caption{Azimuthal anisotropy $v_2$ as a function of
the transverse momentum for charged mesons and baryons
identified in the RICH detector (left)  and neutral kaons and
lambdas identified via secondary vertices in the TPC (right).
Only statistical errors are show.}
\vspace{-10mm}
\label{fig:mesonbaryon}
\end{figure}

Figure \ref{fig:mesonbaryon} shows $v_2$ for identified particles as
a function of transverse momentum. The left panel shows pions,
kaons, and (anti-)protons identified at $p_t<1$ GeV/c via dE/dx in the
TPC, while at higher $p_T$ charged mesons and protons/anti-protons
are identified  track by track in the RICH detector. The right panel
shows neutral kaons and lambdas identified via secondary vertices in
the TPC. 
For all species,
the low momentum behaviour of $v_2$ is very well 
described by the hydrodynamical calculations \cite{hydro}. At higher momenta
this prediction breaks down, baryons and mesons  no longer exhaust
the hydrodynamical limit; but their $v_2$ appears to saturate. Looking
at the combined results, the heavier baryons may
have a slightly higher $v_2$ than the mesons, though the magnitude of
the difference is markedly smaller than predicted in \cite{glvp}.

The transverse momentum dependence of the anti-proton
to proton ratio
together with the anti-lambda/lambda ratio
was suggested \cite{wangpbarp}
as a means to study energy loss of gluons
versus quarks in a pQCD scenario, where the ratio is predicted to
decrease with increasing $p_T$. However, not all observed
anti-protons and protons come from 
fragmentation of jets.  At lower transverse momenta, soft processes and
possibly baryon transport play an  important role.  The transition from soft to
hard contributions to the ratio; i.e., the transverse momentum dependence
of the relative cross sections, is discussed in \cite{vitevqm}. An observed
decrease of the ratio would suggest that the transverse momentum
regime at which perturbative cross sections dominate
has been reached.

\begin{floatingfigure}{90mm}
\includegraphics*[width=90mm]{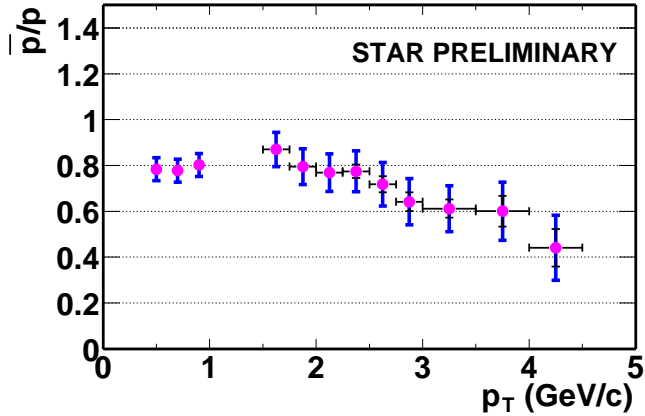}
\vspace{-3mm}
\caption{Anti-proton to proton ratio for 10~\% central
  Au+Au collisions at $\sqrt{s_{NN}}$ = 200 GeV as
a function of transverse momentum. The low $p_T$ particles were 
identified in the TPC while at the high
$p_T$ the RICH detector was used for particle identification.
Statistical and systematic errors are shown.}
\vspace{2 mm}
\label{fig:pbarp}
\end{floatingfigure}

\noindent
Figure \ref{fig:pbarp} shows the ratio of anti-prot\-ons to protons as a
function of  transverse momentum for the 10\% most central
collisions at $\sqrt{s_{NN}}$ = 200 GeV.
The identification of the (anti)protons at $p_T <$
1 GeV was performed track by track via the specific energy loss
dE/dx in the TPC \cite{gene}. Particle
identification at high transverse momentum is obtained using a
statistical analysis of the Cherenkov angle measurements with the 
STAR-RICH detector \cite{brian}. 
The ratio appears to be $p_T$ independent
up to about 2 GeV/c. This momentum independence is
qualitatively in line with earlier measurements at $\sqrt{s_{NN}}$ =
130 GeV \cite{qm01dunlop}, though the absolute ratio is larger
at 200 GeV/c.
The ratio decreases with increasing momentum achieving 0.44+-0.14
(stat + sys. errors) at 4.25 GeV/c.
Together with the evidence from the previous section; i.e., the breakdown
of the hydrodynamic prediction for proton $v_2$, this indicates
pQCD dominance of the cross section in the region $p_T=$3-4 GeV/c.

Recent theoretical papers suggest the use of jets for a tomography
of the medium created in Au+Au collisions. Azimuthal angular correlations
of high transverse momentum track pairs were invoked earlier in this paper
to indicate the existence of jets in Au+Au collisions.
Here we present a new observable: back-to-back azimuthal correlations,
details can be found in  \cite{dave}.
We summarize here the  assumptions, briefly outline the procedure,
and then present the results.

\begin{floatingfigure}{100mm}
\includegraphics*[width=95mm]{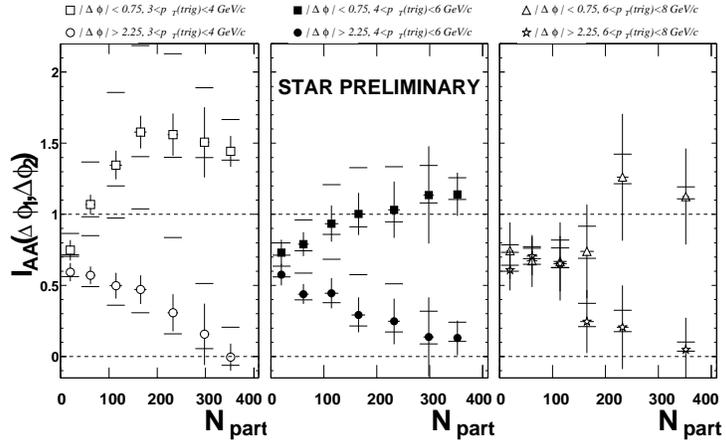}
\vspace{-4mm}
\caption{Ratio $I_{AA}$ (equation \ref{daveeq}), of Au+Au and
p+p for near-side and  back-to-back azimuthal regions  versus
number of participating nucleons. Three trigger particle
thresholds $3<p_T<4$ GeV/c, $4<p_T<6$ GeV/c and
$6<p_T<8$ GeV/c are shown in the left, middle and right panel
respectively.}
\vspace{2mm}
\label{fig:IAA}
\end{floatingfigure}

Because back-to-back jet correlations are not short range in $\eta$,
we cannot separate elliptic flow and jet correlations in this case by
studying different $\Delta \eta$ intervals. 
In the following, we compare Au-Au collisions with proton-proton
collisions taking into account the effects of elliptic flow,
which was determined independently by the reaction plane
me\-thod \cite{v2charged,v2highpt}.


In order to compare Au+Au correlation functions with the  nucleon-nucleon
case, a simple ansatz is suggested ~\cite{dave}.
The correlation function $D(Au+Au)$ for Au+Au is assumed
to be a  superposition of the correlation function $D(p+p)$ and a component
describing the azimuthal anisotropy of the two particles due to the
known $v_2$ in Au+Au  collisions for the respective centrality
class (see \cite{kirill} and above). The explicit functional form used in
the study is:\\

\begin{equation}
D^{\mathrm{AuAu}} = D^{\mathrm{pp}} + B(1+2v_2(p_T^t)v_2(p_T^a) 
\cos(2\Delta
  \phi)),
\label{c2eqn}
\end{equation}

where {\it t} and {\it a} denote the trigger and
associate particle, respectively. The proportionality factor $B$
is determined by fitting the region $0.75<|\Delta \phi|<2.24$, which
is free from jet contributions \cite{dave}.

A quantitative interpretation of figure \ref{fig:azimuthal} is 
obtained by constructing the ratio $I_{AA}$:

\begin{equation}
I_{AA}(\Delta \phi_1,\Delta \phi_2) =
\frac{\int_{\Delta \phi_1}^{\Delta \phi_2} d(\Delta \phi) 
[D^{\mathrm{AuAu}}
- B(1+2v_2^t v_2^a \cos(2 \Delta \phi))]}{\int_{\Delta \phi_1}^{\Delta 
\phi_2} d
(\Delta \phi) D^{\mathrm{pp}}}.
\label{daveeq}
\end{equation}

A ratio of unity implies azimuthal
two-particle correlations in Au+Au collisions are explained by a
simple superposition of p+p collisions plus elliptic flow.

Figure \ref{fig:IAA} shows the ratio $I_{AA}$ for three classes of trigger
particles for near ($|\Delta \phi|<0.75$) and back-to-back
($|\Delta \phi|>2.24$) azimuthal regions.
The near-side ratios are shown square symbols, the back-to-back
ratio as round symbols, the vertical error bars indicate the
statistical error, while the horizontal error bars show the
systematic error obtained by varying the $v_2$ contribution by
+5-20\%.
The near-side ratios increase with the number of
participants (centrality), while the back-to-back ratio decreases
with $N_{part}$.
In the most central Au+Au collisions, the near-side
correlations are similar to those observed in  p+p,
while the back-to-back  correlation strength is consistent with zero.
The near-angle and back-to-back correlation measurements together suggest:
(1) near-side jet properties are approximately unmodified by the
nuclear medium, and (2)
jets  originate from the surface of the collision region. The disappearance of
back-to-back correlations indicates strong medium effects for those jets which
are either produced in the central region or propagate through the central
region.

\begin{floatingfigure}{90mm}
\includegraphics*[width=88mm]{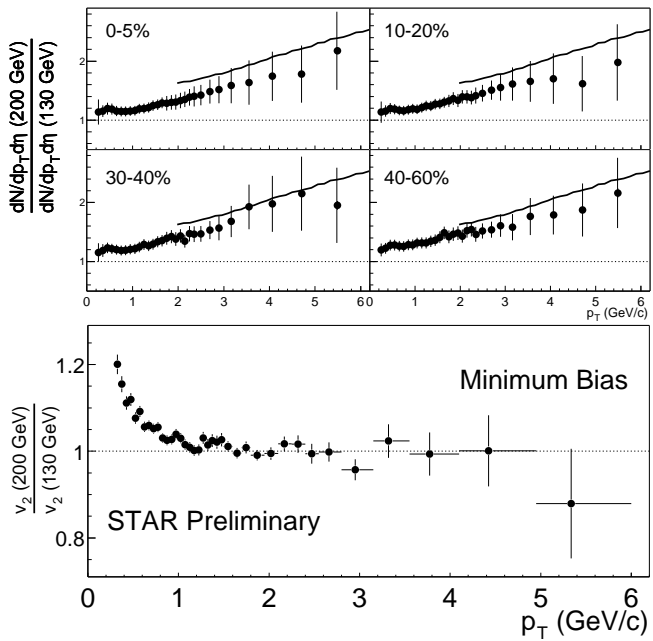}
\vspace{-6mm}
\caption{Upper panel: Ratio of invariant cross sections for charged hadrons
as a function of transverse momentum for Au+Au collisions at
$\sqrt{s_{NN}}$ = 200 GeV and 130 GeV for four
different centralities. The lines depict the
expectations for the $\sqrt{s}$ dependence predicted by pQCD.
Lower panel: Ratio of $v_2$ as  a function of transverse momentum for Au+Au 
collisions at $\sqrt{s_{NN}}$ = 200 GeV and 130 GeV for minimum
bias events.}
\vspace{2mm}
\label{fig:200130gev}
\end{floatingfigure}

\noindent
Additional evidence for this picture is found in 
the $\sqrt{s}$ dependence of had\-ron production and azimuthal
an\-iso\-trop\-ies at high transverse momentum.
The top panel of figure
\ref{fig:200130gev} shows the ratio of invariant cross sections
as a function of $p_T$ for Au+Au collisions at
$\sqrt{s_{NN}}$ = 200 GeV and 130 GeV for four different centralities.
The hadron yield increases by 15-20\% at low $p_T$
for the 200 GeV data ($p_T$ integrated 
yields are dis\-cuss\-ed in \cite{gene}). At a transverse momentum
of 6 GeV, the hadron yield approximately doubles 
from 130 to 200 GeV. The solid line illustrates the $\sqrt{s}$
dependence predicted by pQCD \cite{xnwang3}.
The ratio at high transverse momentum is within the error compatible with
the pQCD predictions. No centrality dependence is observed.

The $\sqrt{s_{NN}}$ dependence of the azimuthal anisotropy $v_2(p_T)$ is
shown in the lower panel of figure \ref{fig:200130gev}. The data for
130 GeV is taken from \cite{v2highpt}.  Elliptic flow
is stronger at $\sqrt{s_{NN}}=200$ GeV for  $p_T <$ 1 GeV/c, see
\cite{kirill}. For transverse momenta above 1 GeV/c, the ratio is close to
unity. The inclusive hadron yields grow with $\sqrt{s_{NN}}$, in line
with pQCD expectations, while elliptic flow is independent of  $\sqrt{s_{NN}}$.
This is consistent with a surface emission picture \cite{bmueller}, in which
the origin of $v_2$ at high $p_T$ is geometric and not dynamical.

The PHOBOS collaboration presented at this conference
(see also \cite{roland} this proceedings) an interesting
finding concerning the scaling behaviour of high $p_T$ hadron production.
Surface effects, as in
the picture discussed in the previous paragraph, scale as
$\langle N_{part} \rangle $. 
In order to study the production
of high transverse momentum hadrons either
$\langle N_{bin} \rangle $ or $\langle
N_{part} \rangle $ can be used to scale $R_{AA}$.
The superposition of nucleon-nucleon hard scattering
in the absence of nuclear effects is constructed by normalizing to
the number of binary collisions. Figure 2 of this paper suggests that
at high transverse momentum the data do not follow this scaling, but
are closer to the $N_{part}$ normalization.

Following the PHOBOS suggestion,
we replot the STAR 130 GeV data in Figure \ref{fig:RaaNpart}.
The ratio $R_{AA}^{N_{part}}$ of invariant
yields measured for Au+Au collisions to the $\langle N_{part} \rangle$ 
scaled NN reference for 6 bins in transverse momentum is
displayed as a function of the number of participants.
The errors bars include both
the statistical and systematic uncertainties from the Au+Au spectra and
the uncertainty on $\langle N_{part} \rangle $. For the tabulated
$\langle N_{part} \rangle $, see \cite{starhighpt}.

The ratios increase significantly with $\langle N_{part} \rangle $ in the
1-3 GeV/c range. A qualitatively different picture is seen in the bottom
right panel, where the ratio is constant for
all centralities. However,
even at high pT
scaling with $\langle N_{part} \rangle $  holds only within the
Au+Au dataset and not with respect to the
nucleon-nucleon reference.

\section{Conclusion}

\begin{floatingfigure}{90mm}
\includegraphics*[width=87mm]{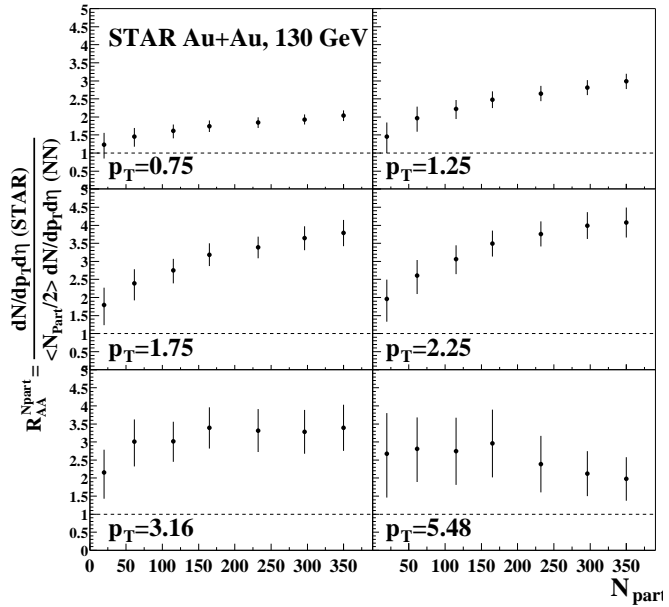}
\vspace{-2mm}
\caption{Ratio $R_{AA}^{N_{part}}$ of
invariant cross sections in AA collisions to the $\langle N_{part} \rangle $
scaled NN reference for 6 bins in transverse momentum as a function of the
number of participants (centrality). The data for 130 GeV are taken from reference
\cite{starhighpt}.}
\vspace{3mm}
\label{fig:RaaNpart}
\end{floatingfigure}
The study of high transverse momentum hadronic
probes yields several observations which must be interpreted in one
consistent picture. 

The experimental evidence for hard scattering lies in the fact that the
two-particle
azimuthal correlations for high $p_T$ hadron\-s show structures which
resemble jet signals, previously established for nucleon-nucleon collisions.
In addition, the anti-proton to proton ratio decreases with
increasing $p_T$, suggesting
that particle production at these momenta is dominated by pQCD processes.

The suppression of high $p_T$ had\-ron inclusive yields in central events
with respect to the binary scaled nucleon-nucleon reference
is consistent with the suppression of the back-to-back jet
signal in the  two-particle azimuthal correlations. This leads to the
following hypothesis for central collisions: The strong absorption in the
bulk matter means that those jets generated in the center of the collision zone
or passing through it are strongly absorbed. The only jets that are
observed are those produced near the surface, directed outwards.


While peripheral collisions appear to be a superposition of nucleon-nucleon 
collisions, this is not the case for central collisions. The question whether
partonic energy loss has been clearly observed, or whether some effects can
be attributed to hadron interactions with the medium, remains to be answered.

\end{document}